\documentclass[twocolumn,showpacs,preprintnumbers,amsmath,amssymb]{revtex4}
\usepackage{graphicx,longtable}
\usepackage{epstopdf,amsthm}
\usepackage{dcolumn}
\usepackage{bm}
\usepackage{color}
\begin{document}
\preprint{APS/123-QED}
\title{Fundamental limit to  Qubit Control with Coherent Field}
\author{Kazuhiro Igeta}
\affiliation{NTT Basic Research Laboratories Nippon Telegraph and Telephone Corporation
\\ 3-1, Morinosato Wakamiya Atsugi-shi, Kanagawa 243-01, Japan
}
\affiliation{Japan Science and Technology Agency, CREST, 5, Sanbancho, Chiyoda-ku, Tokyo, 102-0075, Japan}
\author{Nobuyuki Imoto}
\affiliation{
Graduate School of Engineering Science, Osaka University, 
Toyonaka, Osaka 560-8531, Japan}
\author{Masato Koashi}
\affiliation{Photon Science Center, the University of Tokyo,
2-11-16, Yayoi, Bunkyo-ku, Tokyo 113-8656, Japan}
\date{09102012}
\begin{abstract} 
The ultimate  accuracy as regards
controlling a qubit with a coherent field
is studied in terms of degradation of the fidelity  
by employing  a fully quantum mechanical treatment.
While the fidelity error 
accompanied  by $\pi/2$ pulse control is
shown to be inversely proportional to the average photon number
in a way similar to that revealed by the Gea-Banacloche's results~\cite{GB}
our results  show that the error depends strongly on the initial state of the qubit.
When  the initial state of the qubit
is  in the ground state, the error is about 20 times smaller than
that of the control started from the exited state, no matter how large $N$ is.
This dependency  is explained in the context of an exact quantum mechanical description of the  pulse area theorem. By using the result, the error accumulation tendency of  successive pulse controls is found to be both non-linear and initial state-dependent. 
\end{abstract}

\maketitle
\newcommand\bra[1]{\langle {#1}|}
\newcommand\ket[1]{|{#1} \rangle}
\newcommand\braket[2]{\langle {#1}|{#2}\rangle}
\newcommand{\sinc}{{\rm sinc}\ }
\newcommand{\B}{\Big}
\newcommand\bottom[1]{_{\textrm {\normalsize #1}}}
\section{Introduction}
Nowadays several two-state systems have been proposed as candidates of physical qubits, and the means of single-qubit control and inter-qubit control are also investigated.  Amongst them, controlling matter qubits with coherent electromagnetic field is of particular interest such as quantum-dot qubit control~\cite{Exiton}, spin qubit control~\cite{Edamatsu}, single-atom qubit control~\cite{vanEnk}, ionic qubit control~\cite{CiracZoller}, and Josephson junction qubit control~\cite{JJ} using optical or microwave pulses.    
In practice, we can use the classical field theory which enables us to control a qubit 
perfectly as described by the Bloch rotation according to the pulse area of the
field~\cite{Rabi,Pulse}.

However, classical fields exist only as an infinitely strong intensity limit.
In reality, strong but finite strength coherent fields are available that will
cause control errors because of  their lack of infiniteness.

Gea-Banacloche reported  that the fidelity error of  qubit control is limited by a value inversely proportional to the average photon number of the field $N$~\cite{GB}.
A fidelity error of  ${\pi^2+4}\over 64N$ is obtained for  $\pi/2$ pulse control of
a qubit
and directly reflects the coherent noise inherent to the control field~\cite{GB}.
 Meanwhile, Ozawa considered a quantum mechanical bound of the control field both for single qubit rotation and two-qubit control gates, and derived a universal error bound for the single qubit rotation to be
 $\frac{1}{16(N+1)}$ from 
the uncertainty principle~\cite{Ozawa},
which  is a rigorous bound but not tight enough for specific models
including this case.
As regards the coherent qubit control, 
Ozawa's bound is ascribed to the phase error part
of the coherent noise
~\cite{GBOzawa} 
but cannot give the value of the total error.

In this paper, we formulate the full quantum mechanical interaction between
a pure coherent field and a qubit in the general initial state including mixed states. 
Then the error rates will be shown to  depend on both the field and  the initial state of the qubit~\cite{cleo}.
We break down the result into the first correction to the classical pulse area
theorem by representing the error rates by the order of $1/N$.
We also show the successive control of a $\pi/2$ pulse with a single
$\pi$ pulse and investigate how the error accumulation differs from that in the
classical case.
Furthermore, the pulse area theorem is deduced as the classical limit
of the map by taking the limit $N\rightarrow\infty$.
Here we assume a control field in a good cavity so that the Jaynes-Cummings interaction~\cite{JC}, dominates over the interaction of the field mode with the external field modes.  Our theory is  applicable 
as far as the Rabi frequency 
is much larger than the cavity decay constant, although we neglect  the
cavity decay and deal with the tendency of the control error for large $N$ cases
in this paper.
In Sec.\ II, we derive a rigorous solution for the qubit dynamics controlled by a
coherent field, and obtain a
completely positive trace preserving (CPTP) map expression in terms of Kraus
operators.
In Sec.\ III,  we  derive 
the general form of the error rate using the CPTP map derived 
in Sec.\ II. The result will be rendered into a simple approximation 
by the order of $1/N$ for special cases corresponding to the $\pi/2$ and $\pi$ pulse
controls for comparison with  previous results.
In Sec.\ IV, successive control and error accumulation are examined.
 Sec.\ V concludes the paper with  discussions deriving the maximum error rate 
during the general control of the qubit.

\section{Exact solution of qubit time evolution induced by quantum fields}
\subsection{Qubit evolution caused by coherent field}
A control field and a qubit interact  during the control process
and then physically separate.
This must introduce decoherence into each divided state. 
The coherence of the qubit is preserved only when the control field is classical, or
the infinitely strong coherent field.

A qubit is assumed to be a two-level system whose upper and lower states are
assigned as $\mbox{``1''}$   and the lower as $\mbox{``0''}$, respectively.
Corresponding state vectors, which are denoted respectively as 
$\ket{\mbox{1}}$ and $\ket{\mbox{0}}$, are written using the parameters
$\phi$ and $\theta_{0}$\, $(0 \le \phi,\theta_{0} \le 2\pi) $ as 
\begin{eqnarray}
\ket{qubit}&
\equiv
&\cos\theta_{0}\ket{\mbox{0}}+\mathrm{e}^{i\phi}\sin\theta_{0}
                                          \ket{\mbox{1}}\\
         &=&\mathrm{e}^{i\phi/2}\hat P(\phi)
                         \left(\begin{array}{c}
                          \sin\theta_{0}\\
                          \cos\theta_{0}
                        \end{array}\right)\bottom{,} 
                        \label{eq:qubit}
\end{eqnarray}
where
\begin{equation}
\hat P(\phi)\equiv\left(\begin{array}{cc}
               \mathrm{e}^{i\phi/2}&0\\
                 0&\mathrm{e}^{-i\phi/2}
                 \end{array}\right)\bottom{,} 
\end{equation}
and the representation basis is
\begin{equation}
\ket{\mbox{1}}\equiv\left( \begin{array}{c}
                          1\\0
                         \end{array}
\right)\bottom{,} 
\phantom{000}
\ket{\mbox{0}}\equiv\left( \begin{array}{c}
                          0\\1
                         \end{array}
\right)\bottom{,} 
\label{eq:unitvector} 
\end{equation}
throughout this paper.
$\cos^{2}{\theta_{0}}$ and $\sin^{2}{\theta_{0}}$ are the probabilities of taking the values $\mbox{``0''}$ 
and $\mbox{``1''}$, respectively, when the qubit is measured in basis (\ref{eq:unitvector}).

In our model, the qubit is controlled by a single-mode initially-pure state field
via a fully quantum mechanical Jaynes-Cummings interaction without detuning~\cite{JC}.
Let the initial total density operator be
\begin{equation}
{\hat\rho}(0) = {\hat\rho}_{\rm f}(0) 
\otimes {\hat\rho}_{\rm q}(0)\bottom{,} 
\end{equation}
where ${\hat\rho}_{\rm f}(0), {\hat\rho}_{\rm q}(0)$ are density
operators for the field and the qubit, respectively.
The density operator after  interaction time $t$ is expressed as
\begin{equation}
{\hat\rho}(t) =\hat{U}{\hat\rho}(0)\hat{U}\bottom{,} 
\label{Eq:rho}
\end{equation}
where $\hat{U}$ is the unitary operator of the evolution.

We assume a non-detuning case, namely, the total Hamiltonian is expressed as
\begin{equation}
{\hat{\cal H}} = \hbar\omega\left({\hat{a}^\dagger \hat{a} } +{\hat\sigma_z}/2 \right) +{\hat{\cal H}}_{I}\, .
\end{equation}
where 
\[
\hat\sigma_z=
\left(
\begin{array}{cc} 
1&0\\
0 & -1
\end{array}
\right)_{\textrm {\normalsize , }}
\]
is 
the Pauli matrix representing population inversion and $\hat{a}^{\dagger},\hat{a}$ are the creation and annihilation operators of the field,
$\hat{\cal H}_{I} $ is the Jaynes-Cummings interaction Hamiltonian,
which is naturally deduced from a linear interaction model with a bosonic
field after taking the rotating wave approximation~\cite{JC},
\begin{equation}
{\hat{\cal H}}_{I} = \hbar g \left({\hat{a}}{\hat\sigma}_{+} 
+ \hat{a}^{\dagger}{\hat\sigma_{-}}\right)\bottom{,} 
\end{equation}
where $g$ is the parameter for 
the interaction strength determined by the qubit dipole moment
and the local field strength, which may be modulated by a cavity, 
and
${\hat\sigma}_{+}$ and ${\hat\sigma_{-}}$ are the elevation
 operators for the qubit,
which are represented by the basis defined
by Eq.~(\ref{eq:unitvector}) as
\begin{eqnarray}
{\hat\sigma}_{+}&=&
\ket{\mbox{1}}\bra{\mbox{0}}=
\left(
\begin{array}{cc} 
0&1\\
0 & 0
\end{array}
\right)\nonumber
\\
{\hat\sigma_{-}}&=& 
\ket{\mbox{0}}\bra{\mbox{1}}=
\left(
\begin{array}{cc} 
0& 0\\
1 & 0
\end{array}
\right)\bottom{.} 
\end{eqnarray}
Using the commutation relations for field and spin operators, 
the evolution operator $\hat{U}$ 
under  interaction picture is expressed as a function of $\kappa\equiv gt$ as
\begin{eqnarray}
{\hat U}(\kappa)&=& \exp\left[-i \kappa \left(\hat{a}{\hat\sigma}_{+} 
+ \hat{a}^{\dagger}{\hat\sigma_{-}}\right)\right]\nonumber\\
&=& \sum_{l=0}^{\infty}{1 \over l!}({\kappa \over
i})^{l}(\hat{a}{\hat\sigma}_{+} + \hat{a}^{\dagger}\hat\sigma_{-})^{l}\nonumber\\
&=&\sum_{l=0}^{\infty}{1 \over (2l)!}({\kappa \over i})^{2l}
\{ ({1 \over {2l+1}}{\kappa \over i})(\hat{a}{\hat\sigma}_{+} 
+ \hat{a}^{\dagger}{\hat\sigma_{-}})^{2l+1}
\nonumber\\&&
+(\hat{a}{\hat\sigma}_{+}+\hat{a}^{\dagger}{\hat\sigma_{-}})^{2l}\}
\nonumber\\
&=&\sum_{l=0}^{\infty}{1 \over (2l)!}({\kappa \over i})^{2l}
\left[({1 \over {2l+1}}{\kappa \over i})
\{(\hat{n}+1)^{l}\hat{a}{\hat\sigma}_{+} +
\hat{n}^{l}\hat{a}^{\dagger}{\hat\sigma_{-}}\}
\right.\nonumber\\&&\left.
+\{{\hat\sigma_{-}}{\hat\sigma}_{+}\hat{n}^{l}+
{\hat\sigma}_{+}{\hat\sigma_{-}}(\hat{n}+1)^{l}\}\right]\nonumber\\
&=& 
\left(
\begin{array}{cc} 
\cos{\kappa\sqrt{\hat{n}+1}} 
& -i\kappa(\sinc\kappa\sqrt{\hat{n}+1})\hat{a}\\
 -i\kappa(\sinc\kappa\sqrt{\hat{n}})\hat{a}^{\dagger}
 & \cos\kappa\sqrt{\hat{n}}
\end{array}
\right)\bottom{,} 
\label{Eq:U}
\end{eqnarray}
where $\hat n\equiv {\hat{a}}^{\dagger} {\hat{a}}$,
\begin{equation}
 \sinc x\equiv \left\{ \begin{array}{cc} 
                         \sin x/\,x & x \neq 0\\
                                  1 & x=0\, .
                      \end{array}\right.
\end{equation}

First, we assume that the field is in a general pure state (not necessarily in 
a coherent state).
Then, ${\hat\rho}_{f}(0)$ is expressed by
photon number expansion as
\begin{equation}
{\hat\rho}_f (0)=\sum_{m=0}^{\infty}\sum_{l=0}^{\infty}
C_m\ket{m}\bra{l}{C_l^{*}}\bottom{,} \label{Eq:Expansion}
\end{equation}
where $\ket{m},\bra{l}$ are the number state vectors of photon numbers
$m$ and $l$, respectively, and $C_m,C_l^{*} \in {\cal C}$ are the coefficients that
correspond to the photon numbers $m$ and $l$, respectively.
Using Eq.~(\ref{Eq:Expansion}), the total density operator after evolution is written as
\begin{eqnarray}
{\hat\rho}(t)&=&
\hat{U}(\kappa){\hat\rho}_{\rm f}(0) \otimes {\hat\rho}_{\rm q}(0)
\hat{U}^{\dagger}(\kappa)\nonumber\\
&=& \sum_{m=0}^{\infty} \sum_{l=0}^{\infty}
\hat{U}(\kappa)\ket{m}C_m{\hat\rho}_{\rm q}(0)C_l^{*}\bra{l}
\hat{U}^{\dagger}(\kappa)\bottom{.}\label{Eq:RhoU}
\end{eqnarray}
Thus, the density operator of a qubit is obtained in a completely positive trace preservation (CPTP) map as
\begin{eqnarray}
{\rm Tr}_{\rm f} \{{\hat\rho}(t)\}
&=&\sum_{n=0}^{\infty}\bra{n}{\hat\rho}(t)\ket{n}\nonumber\\
&=&\sum_{n=0}^{\infty}{\hat M}_{n}{\hat\rho}_{q}(0){{\hat M}_{n\,.}^{\dagger}}
\label{Eq:Reduced}
\end{eqnarray}
where the Kraus operator ${\hat M}_{n}$ is described as
\begin{widetext}
\begin{eqnarray}
{\hat M}_{n}&\equiv&\sum_{m=0}^{\infty}
\bra{n}\hat{U}(\kappa)\ket{m}C_m\nonumber\\
&=&
\sum_{m=0}^{\infty} C_m \left(
  \begin{array}{cc} 
  \delta_{n,m}\cos{\kappa\sqrt{m+1}} &
  -i\kappa \sqrt{m} \delta_{n+1,m}\sinc\kappa\sqrt{m}\\
  -i\kappa\sqrt{m+1}\delta_{n,m+1}\sinc\kappa\sqrt{m+1}&
  \delta_{n,m}\cos{\kappa\sqrt{m}}
\end{array}\right)\nonumber\\
&=& \left(
\begin{array}{cc} 
C_n \cos{\kappa\sqrt{n+1}} &
-iC_{n+1}\sin\kappa\sqrt{n+1}\\
-iC_{n-1}\sin\kappa\sqrt{n} & C_n\cos{\kappa\sqrt{n}}
\end{array}
\right)\bottom{,}  \label{Eq:Un}
\end{eqnarray}
\end{widetext}
where $C_{-1}\equiv 0$ is added for the definition of the field coefficient.

Eqs.~(\ref{Eq:Reduced}) and (\ref{Eq:Un}) allow us to 
calculate any qubit state including a fully mixed state evolved by any initially pure quantum field.
As an  example, qubit states after the 
$\pi/2$-pulse control with  a variety of control field types, i.e., a number state, a classical state, or a coherent state, are shown 
in Table~\ref{table1}.
\begin{table*}
\begin{tabular}{c|c}\hline
Field type&${\hat\rho}_q (t)$\\
\hline\hline 
&\\
 \begin{tabular}{c} 
   Classical state \\
\end{tabular}
&
${1\over 2}\left(
\begin{array}{cc}
     1 & 1\\
     1 & 1
\end{array}
\right)$  \\
&Ideal Control 
\\
\hline
&\\
Number state
 &
 ${1\over 2}\left(
\begin{array}{cc} 
   1& 0 \\
   0 &1
\end{array}
\right)$
\\ 
& Qubit becomes completely mixed.
\\
\hline
&\\
Coherent state &
$
{\displaystyle \sum_{n=0}^\infty}
|C_{n}|^2 \left(\begin{array}{cc}
\sin^2\left(\frac{\pi}{4}\sqrt{\frac{n}{N}}\right)  &
 \sqrt{\frac{N}{n+1}}\cos\left(\frac{\pi}{4}\sqrt{\frac{n}{N}}\right)
 \sin\left(\frac{\pi}{4}\sqrt{\frac{n+1}{N}}\right) 
\\
\sqrt{\frac{N}{n+1}}\cos\left(\frac{\pi}{4}\sqrt{\frac{n}{N}}\right) 
\sin\left(\frac{\pi}{4}\sqrt{\frac{n+1}{N}}\right) &
\cos^2\left(\frac{\pi}{4}\sqrt{\frac{n}{N}}\right)
\end{array} \right)
$
\\
\\
\hline
\end{tabular}
\caption{
 Qubit states controlled from the ground  state $\ket{0}$ to an evenly occupied state
$(\ket{0}+\ket{1})/\sqrt{2}$ are shown.
The number state field makes the qubit completely mixed while the 
classical state (strong coherent state limit)
maintains its purity by employing a $\pi/2$ pulse. The coherent state with finite strength is expressed by the sum of
each photon number component  where $\kappa$ is set  as $\pi/(4\sqrt{N})$
to correspond to the pulse area $\pi/2$ .
The coherent sate $\pi/2$ pulse cannot give the ideal control as far as $N$ is finite. }
\label{table1}
\end{table*}
\subsection{Pulse Area Theorem as Classical Limit}
When we assume that the  control field is a coherent state with strong intensity,
 i.e.,  the average photon number $N$ is very large, 
the pulse area theorem states that the CPTP map (\ref{Eq:Reduced})  should  be well 
approximated by the classical Bloch rotation
\begin{equation}
R(\vartheta)\hat{\rho}_q(0)R^\dagger(\vartheta)\bottom{,} 
\nonumber
\end{equation}
where
\begin{equation}
R(\vartheta)\equiv 
\left(\begin{array}{cc} 
\cos\vartheta &
\sin\vartheta\\
-\sin \vartheta&
\cos\vartheta
\end{array}\right)\bottom{,} 
 \label{Eq:Bloch}
\end{equation}
and $ \vartheta\equiv\kappa\sqrt{N}$ is half of the pulse area of the control field.
In this section, 
we examine the $N\rightarrow \infty$ limit of (\ref{Eq:Reduced}) to prove the theorem,
which is further refined in Sec.\ III to obtain an asymptotic expression in terms of $1/N$.
Noting that the field coefficients of coherent states satisfy the relation $C_{n+1}/C_n=\sqrt{N/(n+1)}\mathrm{e}^{i\varphi}$ 
and  $|C_{n}|^2={\rm e}^{-N}N^{n}/ n!$ for $ n=0,1,2,\cdots$, 
the contribution of the $n$ photon number component of the Kraus operators are written 
as, 
\begin{eqnarray}
{\hat M}_{n}
&=& 
C_n \hat P\bigg(\varphi-\frac{\pi}{2}\bigg) 
{\hat m}\bigg(\frac{n-N}{N}\bigg)
 \hat P\bigg(-\varphi+\frac{\pi}{2}\bigg)
\bottom{,} \nonumber\\
\label{Eq:Rotate}
\end{eqnarray}
where
\begin{eqnarray}
{\hat m}(x) 
\equiv \left(\begin{array}{cc} 
\cos \bigg(\vartheta\sqrt{1+x+\frac{1}{N}}\bigg) &
\frac{\sin \bigg(\vartheta\sqrt{1+x+\frac{1}{N}}\bigg)}
{\sqrt{1+x+\frac{1}{N}}}\\
-\sqrt{1+x}\sin\Big(\vartheta\sqrt{1+x}\Big) &
\cos\Big(\vartheta\sqrt{1+x}\Big)
\end{array}\right)_{\textrm {\normalsize .}}
 \nonumber\\
\label{Eq:Defm}
\end{eqnarray}
Choosing $\varphi=\pi/2$  to eliminate the phase factor in Eq.(\ref{Eq:Rotate}),
we obtain from (\ref{Eq:Reduced}), (\ref{Eq:Rotate}) and (\ref{Eq:Defm}) that
\begin{equation}
\lim_{N\rightarrow \infty}{\rm Tr}_{\rm f} \{{\hat\rho}(t)\}
=\lim_{N\rightarrow \infty}
\sum_{n=0}^{\infty}{|C_n|}^{2}
{\hat m}(x){\hat\rho}_{q}(0){{\hat m}(x)^{\dagger}}
\B|_{x=\frac{n-N}{N}\,\textrm{\normalsize .}}
\label{Eq:Area1}
\end{equation}
In order to calculate (\ref{Eq:Area1}),
let us define $\mu_{k}$ which corresponds to  the $k$-th central moment of $n/N$ as
\begin{eqnarray}
\mu_{k}&\equiv&
\sum_{n=0}^{\infty}{|C_n|}^{2} x^k \Big|_{x=\frac{n-N}{N}} \nonumber\\
&=&
 \sum_{n=0}^{\infty} \mathrm{e}^{-N}  \frac{N^{n}}{n!} \bigg(\frac{n-N}{N}\bigg)^k_\textrm{\normalsize .}
\label{Eq:mu}
\end{eqnarray}
Obviously,
\begin{eqnarray}
\mu_0&=&1,\label{Eq:mu0}\\
\mu_1&=&0\, ,\label{Eq:mu1}
\end{eqnarray}
and for $k\ge 2$,
\begin{eqnarray}
\mu_{k}&=&\mathrm{e}^{-N}  \sum_{n=0}^{\infty}\frac{N^{n-k}}{n!}\bigg\{\ n (n-N)^{k-1}
-{N}(n-N)^{k-1}\bigg\}\nonumber\\
&=&\mathrm{e}^{-N}  \sum_{n=0}^{\infty}\frac{N^{n-k+1}}{n!} \Big\{
(n+1-N)^{k-1}-(n-N)^{k-1}\Big\}\nonumber\\
&=&  \sum_{i=1}^{k-1}
\frac{(k-1)!}{i!(k-1-i)!}\frac{\mu_{k-1-i}}{N^{i}}_{\textrm {\normalsize .}}
\label{Eq:muk}
\end{eqnarray}
Using (\ref{Eq:mu0})-(\ref{Eq:muk}), 
we find that
\begin{eqnarray}
\mu_{k}&\le&\frac{1}{N},
\phantom{0} \mathrm{for}\phantom{0}    k\ge 1.
\label{Eq:xk}
\end{eqnarray}
Eqs.~(\ref{Eq:mu0}), (\ref{Eq:mu1}) 
and (\ref{Eq:xk}) mean that only the $0$-th order term of $x$ contributes
to r.h.s. of (\ref{Eq:Area1}) when $ {\hat m}(x){\hat\rho}_{q}(0){\hat m}(x)^{\dagger}$
is expanded as a power series of $x$.
Therefore, the sum in (\ref{Eq:Area1}) is carried out as
\begin{eqnarray}
\lim_{N\rightarrow \infty}{\rm Tr}_{\rm f} \{{\hat\rho}(t)\}
&=&\lim_{N\rightarrow \infty}
{\hat m}(0){\hat\rho}_{q}(0){\hat m}(0)^{\dagger}
\nonumber\\
&=& R(\vartheta){\hat\rho}_{q}(0)R^\dagger (\vartheta)_{\textrm {\normalsize ,}}
\label{Eq:Area2}
\end{eqnarray}
which  is turned out to be a quantum mechanical proof of the pulse area theorem.
\section{Fidelity of $\pi/2$-pulse Control with Field with Finite Strength}
When the quantum nature of the control field is not negligible,
the control of rotation from $\theta_0$ to $\theta_{0} +\theta$ will be imperfect.
To estimate the imperfection of the control, 
we define the error rate by using the fidelity $\cal F$ as below after 
the preceding study~\cite{GB}.
\begin{eqnarray}
P &\equiv& 1-{\cal F}^2\nonumber\\
                   &\equiv&1-\bra{\psi}{\hat\rho}_{q}(t)\ket{\psi}\nonumber\\
                   &=& \bra{\psi^{\perp}}{\hat\rho}_{q}(t)\ket{\psi^{\perp}}\bottom{,} 
\label{eq:Fidelity}
\end{eqnarray}
where $\ket{\psi^{\perp}}$ is the state orthogonal to the target state$\ket{\psi}\bottom{.} $
Using Eqs.~(\ref{Eq:Reduced}), (\ref{Eq:Un}) and (\ref{Eq:Defm}),
 the error rate for the general control from $R(\theta_0)\ket{0}$ to {$R(\theta_0+\theta)\ket{0}$} can be described as
\begin{widetext}
\begin{eqnarray}
P(\theta_0,\theta)&\equiv& \bra{\psi^{\perp}}\sum_{n=1}^{\infty}{\hat M}_{n}{\hat\rho}_{q}(0){\hat M}_{n}^{\dagger}
\ket{\psi^{\perp}}\nonumber\\
&=&\bra{1}R(-\theta_0-\theta)\sum_{n=0}^{\infty}
{\hat M}_{n}
R(\theta_0)\ket{0}\bra{0}R(-\theta_0)
{\hat M}_{n}^{\dagger}
R(\theta_0+\theta)\ket{1}\nonumber\\
&=&\sum_{n=0}^{\infty}|C_{n}|^2
|\sin{\theta_0}\{m^{11} \cos({\theta_0 +\theta})-m^{01}\sin({\theta_0 +\theta})\}
\nonumber\\ && 
\phantom{000000}
+\cos{\theta_0}\{m^{10}\cos({\theta_0 +\theta})
-m^{00} \sin({\theta_0 +\theta})\}|^2\bottom{,} 
\label{eq:Ptheta}
\end{eqnarray}
where $m^{ij}\equiv \bra{i}{\hat m}(x)\ket{j}$ as shown in (\ref{Eq:Defm}).
Focusing on the $\pi/2$ pulse controls starting  from $\ket{0}$ and 
$\ket{1}$,  the error rates are obtained as below.
\begin{eqnarray}   
P^{+}&\equiv&P(\pi/2,\pi/4)\nonumber\\
&=&\sum_{n=0}^{\infty}|C_{n}|^2
\B\{\frac{\frac{n}{N}\sin^2 (\kappa\sqrt{n})+\cos^2 (\kappa\sqrt{n+1})}{2}-
\sqrt\frac{n}{N}\sin (\kappa\sqrt{n})\cos (\kappa\sqrt{n+1})\B\}
\nonumber\\
&=&\frac{1}{2}-\sum_{n=1}^{\infty}|C_{n}|^2
\sqrt\frac{n}{N}\sin (\kappa\sqrt{n})\cos (\kappa\sqrt{n+1})
\nonumber\\
\label{eq:Pfup}
\\
P^{-}&\equiv&P(0,\pi/4)\nonumber\\
&=&\sum_{n=0}^{\infty}|C_{n}|^2
\B\{\frac{\frac{N}{n+1}\sin^2 (\kappa\sqrt{n+1})+\cos^2 (\kappa\sqrt{n})}{2}
-\sqrt{\frac{N}{n+1}}\sin (\kappa\sqrt{n+1})\cos (\kappa\sqrt{n})\B\}
\nonumber\\
&=&\frac{1}{2}-\sum_{n=0}^{\infty}|C_{n}|^2
\sqrt{\frac{N}{n+1}}\sin (\kappa\sqrt{n+1})\cos (\kappa\sqrt{n})
\nonumber\\
&=&
\frac{1}{2}-\sum_{n=1}^{\infty}|C_{n}|^2
\sqrt\frac{n}{N}\sin (\kappa\sqrt{n})\cos (\kappa\sqrt{n-1})
\nonumber\\
\label{eq:Pfdown}
\end{eqnarray}
\end{widetext}
{
For a fixed $\vartheta(\equiv\kappa\sqrt{N})$,
 Eqs.~(\ref{eq:Pfup})  and (\ref{eq:Pfdown})  can be written as
a single formula as
\begin{equation}
P^{\pm}=\frac{1}{2} -\sum_{n=1}^{\infty}
|C_{n}|^2 f^{\pm} \left(\frac{n-N}{N}\right)
_{\textrm {\normalsize ,}
}
\label{Eq:Pfupdown}
\end{equation}
where
\begin{eqnarray}
f^{\pm} (x)&\equiv& 
\sqrt{1+x}\sin  (\vartheta\sqrt{1+x})\cos \Bigg(\vartheta\sqrt{1+x\pm\frac{1}{N}}\Bigg)_{\textrm {\normalsize .}}
\nonumber\\
\label{Eq:fx}
\end{eqnarray} 
In order to obtain $P^{\pm}$, let us expand $f^{\pm}(x)$  in powers of $x(=\frac{n-N}{N})$ 
as
\begin{equation}
f^{\pm}(x)=f_{0}^{\pm}+x f_{1}^{\pm}+x^2 f_{2}^{\pm}+\cdots.
\label{Eq:fexpansion}
\end{equation}
From (\ref{Eq:mu}), (\ref{Eq:mu0}) and
\begin{eqnarray} 
\mu_{2}&=&{N^{-1}}_{\textrm {\normalsize ,}}\\
\mu_{k}&\le&o\big(N^{-1}\big)_{\textrm {\normalsize ,}}\phantom{0} \mathrm{for}\phantom{0}    k\ge 3,
\end{eqnarray} 
derived from (\ref{Eq:muk}), 
we have the expression
\begin{eqnarray} 
\sum_{n=0}^{\infty}|C_{n}|^2 f^{\pm} \left(\frac{n-N}{N}\right)
&=&\mu_{0}f_{0}^{\pm}+\mu_{1} f_{1}^{\pm}+\mu_{2} f_{2}^{\pm}+\cdots\nonumber\\
&=&f_{0}^{\pm}+\frac{{f_{2}}^{\pm}}{N}+O(N^{-2})_{\textrm {\normalsize .}}
\label{Eq:fsum}
\end{eqnarray} 
Since, the expansion of  (\ref{Eq:fx}) gives
\begin{eqnarray}
f_{0}^{\pm}&=&\frac{\sin2\vartheta}{2}\mp\frac{\vartheta}{2N}\sin^2 2\vartheta +O(N^{-2}),\\
f_{2}^{\pm}&=&\frac{\vartheta}{8}\cos 2\vartheta-\frac{1}{16}(4\vartheta^2+1)\sin 2\vartheta +O(N^{-1})_{\textrm {\normalsize ,}}
\end{eqnarray}
we obtain from (\ref{Eq:Pfupdown}) and (\ref{Eq:fsum}) as
\begin{eqnarray} 
P^{\pm}&=&\frac{1- \sin2\vartheta}{2}+\frac{1}{16N} \{ (1+4\vartheta^2)\sin2\vartheta 
-2\vartheta\cos2\vartheta\nonumber\\&&
\pm4\vartheta(1-\cos2\vartheta)  \}+O\big(N^{-2}\big)_{\textrm {\normalsize .}}		
\label{Eq:Ptheta}
\end{eqnarray} 
Taking large $N$ limit of  (\ref{Eq:Ptheta}) as
\begin{eqnarray}
\lim_{N\rightarrow \infty}P^{\pm}
&=&
\frac{1-\sin2\vartheta}{2}
_{\textrm {\normalsize ,}}
\label{Eq:Pinf}
\end{eqnarray}
$\vartheta=\pi/4(=\theta)$
is found to be the best for  $\pi/2$ pulse control at the classical limit.

For finite $N$, 
$\vartheta=\pi/4$ gives
\begin{eqnarray}
\label{Eq:Ppmc}
P^{\pm} &=&
\frac{(\pi\pm2)^{2}}{64N}
+O\big(N^{-2} \big)_{\textrm {\normalsize .}}
\end{eqnarray}
However, 
(\ref{Eq:Ppmc}) may not give the minimum error rates, because
 $\pi/4$ is no longer the optimal value  for finite $N$.
The optimal $\vartheta$ { to make $P^{\pm}$ minimum} are obtained by solving 
\begin{equation}
\frac{dP^{\pm}}{d\vartheta}=0,
\end{equation}
as
\begin{equation}
\vartheta^{\pm}=\frac{\pi}{4}-\frac{1}{N}\B(\frac{3\pi}{32} \pm\frac{\pi+ 2}{16}\B)+O(N^{-2})_{\textrm {\normalsize ,}}
\label{Eq:thetaopt}
\end{equation}
where $\vartheta^{\pm}$ is the optimal  $\vartheta$ for $P^{\pm}$.
The  differences of $\vartheta^{\pm}$ from $\pi/4$ in (\ref{Eq:thetaopt})  contribute to $P^{\pm}$
in the $N^{-2}$ order.
Therefore, (\ref{Eq:Ppmc})  { is approved as the expression for minimum $P^{\pm}$}.  
In other words,  $O\big(N^{-1}\big)$ tolerance is allowed for setting 
$\vartheta$ when $N$ is large  enough.

The above argument is solely aimed at optimizing the fidelity, and the choice
of $\vartheta$ in (\ref{Eq:thetaopt}) gives the optimal fidelity up to the order of $O(1/N^{2})$.
As we will soon see, this choice results in a bias in $\langle\hat{\sigma_z}\rangle$ of
order $O(1/N)$. Hence, there may be cases where another choice of $\vartheta$
is preferred, which suppresses the bias in $\langle\hat{\sigma_z}\rangle$ to the order
of  $O(1/N^{2})$ while achieving the optimal fidelity up to the order of 
$O(1/N)$.
Instead of $P^{\pm}$, we will consider the deviation of the diagonal elements  of the qubit density operators from the target value $1/2$. 
Using (\ref{Eq:Reduced}) and  (\ref{Eq:Un}), the deviations, $\Delta^{\pm}$, 
are obtained as
\begin{eqnarray}
\label{LevelError}
\Delta^{\pm}&=&\left|\frac{1}{2}\sum_{n=0}^{\infty}|C_{n}|^2 \cos\Bigg( 2\vartheta\sqrt{\frac{n+\frac{1\pm1}{2}}{N}}\Bigg)\right|
_{\textrm {\normalsize .}}
\end{eqnarray}
When $\vartheta=\pi/4$, the leading terms of  $\Delta^{\pm}$ are  $\frac{(2\pm1)\pi}{32N}$,
which vanish by setting  $\vartheta=\tilde{\vartheta}^{\pm}$, where
\begin{equation}
\label{Eq:thetaopt2}
\tilde{\vartheta}^{\pm}=\frac{\pi}{4}-\frac{\pi}{32N} (1\pm2)_{\textrm {\normalsize .}}
\end{equation}
Although (\ref{Eq:Ppmc}) is an asymptote that the actual error rate of 
the $\pi/2$ pulse control converges only at $N\rightarrow\infty$,
Fig.~\ref{fig:ERRORS} shows that it can be a good approximation for 
even in the range $N\lesssim100$. 
Fig.~\ref{fig:ERRORS} also makes it clear
how our results are compared 
to the previous 
results~\cite{GB,GBOzawa}.
All the asymptotic curves in Fig.~\ref{fig:ERRORS}
 are of the order $1/N$ but our curve is not unique 
 but varies greatly up to about 20 times depending  on the initial state of the qubit.
The curves corresponding to $P^{\pm}$ are simply selected as the typical condition
but their average value happens to be the same as the semiclassical result~\cite{GB}.
It is interesting that the error rate starting from the ground state $P^{-}$ is so small
that the curve stays lower than the curve $\frac{1}{ {16(N+1)}}$ derived from the uncertainty principle. 
This is not a contradiction because the curve $1\over {16(N+1)}$ is the lowest limit of the
worst case for the control~\cite{Ozawa} and
the curve for $P^{+}$ in Fig.~\ref{fig:ERRORS} locates above it apparently.
  
\begin{figure}[t]
\includegraphics[scale=0.3]{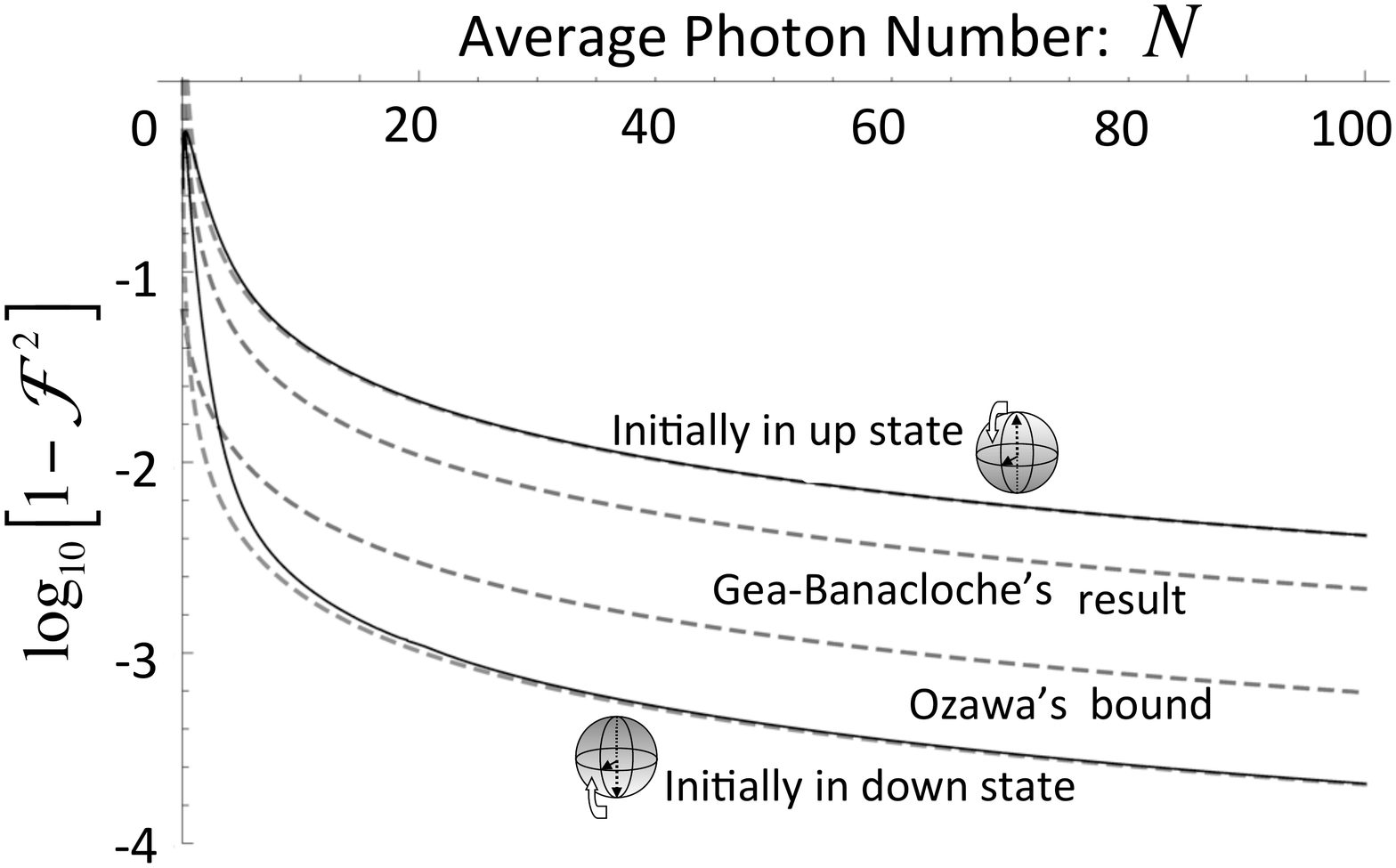}
\caption{
 The solid lines are exact error rates ($1-{\rm fidelity}^2$)  plotted to the average photon
 number of the field $N$ on a $\log_{10}$ scale for the initially ``0"(down) qubit case and  the initially  ``1" (up) state qubit case (from bottom to top).
Both show good asymptotic agreement even for a small $N$ of $\sim 100$,
${\pi^2+4}\over 64N$ (Gea-Banacloche's result), $1\over {16(N+1)}$ (Ozawa's result),
from top to bottom.}
\label{fig:ERRORS}   
\end{figure}
\section{Error Rate and Decoherence after Successive Flipping Controls}
In Sec. III, we quantified the error rates in the qubit controlled
by a coherent field, which  depend not only on the strength of the control field but
also on the the initial condition of the qubit. 
Since (\ref{Eq:Reduced}) allows any qubit state including a mixed state as its initial state,
we can calculate the error rates of successive controls.
Here, we consider a simple case, i.e., successive $\pi/2$ pulse controls
to the qubit initially in the ground state or in the excited state.

Let $P_{\rm d}^{-}$ denotes
the error rate for the successive $\pi/2$ pulse controls that intend to make
$\theta=0\rightarrow\pi/4$ ($\pi/2$ pulse) and $\theta=\pi/4\rightarrow\pi/2$ ($\pi/2$ pulse).
If we set 
$\vartheta=\tilde{\vartheta}^{\pm}$
instead of $\pi/4$ for each control,
$\langle\hat{\sigma_z}\rangle$ vanishes up to 1/N order.
Therefore, the density operator after the first control 
$\theta=0\rightarrow\pi/4$   
can be approximated  as the
statistical mixture up to $O(1/N)$ as
\begin{eqnarray}
{\rm Tr}_{\rm f} \{{\hat\rho}(t)\}=\{1-P(0,\pi/4)\}\ket{\pi/4}\bra{\pi/4}&&\nonumber\\
+P(0,\pi/4)\ket{3\pi/4}\bra{3\pi/4} +O(1/N^{2}),&&
\label{Eq:mixed}
\end{eqnarray}
where $\displaystyle \ket{\pi/4}\equiv\frac{\ket{0}+\ket{1}}{\sqrt{2}}$ is the target state and
$\displaystyle \ket{3\pi/4}\equiv\frac{\ket{0}-\ket{1}}{\sqrt{2}}$ is its orthogonal state.
The total error rate after the second $\pi/2$ pulse control is then written simply as 
}
\begin{eqnarray}
P_{\rm d}^{-}&=&P(\pi/4,\pi/4)\{1-P(0,\pi/4)\}\nonumber\\
&&+\{1-P(3\pi/4,\pi/4)\}P(0,\pi/4)\bottom{.} 
\end{eqnarray}
We will calculate $P_{\rm d}^{-}$ to the order of $O(1/N)$. $P(0,\pi/4)$ was obtained in Eq. (\ref{Eq:Ppmc})
 in the previous section as
\begin{equation}
P(0,\pi/4)=P^{-}=\frac{(\pi-2)^2}{64N}+O(1/N^2)\bottom{.} 
\end{equation}
$P(\pi/4,\pi/4)$ is also obtained using (\ref{Eq:Defm}) and (\ref{eq:Ptheta}) as
\begin{eqnarray}
P(\pi/4,\pi/4)&=&\frac{1}{2}\sum_{n=0}^{\infty}|C_{n}|^2
|m^{00}+m^{01}|^2\nonumber\\
&=&\frac{1}{2}\sum_{n=0}^{\infty}|C_{n}|^2
\bigg\{1-\frac{1}{2}\cos\B(\frac{\pi}{2}\sqrt{1+x}\B)\nonumber\\
&&\phantom{000000}-\sqrt{1+x}\sin\B(\frac{\pi}{2}\sqrt{1+x}\B)\bigg\}\nonumber\\
&=&\frac{(\pi+2)^2}{64N}+O(1/N^2)\bottom{.} 
\end{eqnarray}
Finally, $P_{\rm d}^{-}$ is expressed by  the simple sum of the error rates of 
the successive $\pi/2$ pulse controls up to the order of $N^{-1}$ as 
\begin{eqnarray}
P_{\rm d}^{-} &\sim& P(0,\pi/4)+P(\pi/4,\pi/4)\nonumber\\
&\sim&\frac{\pi^2+4}{32N}\bottom{.} 
\label{eq:Accum1}
\end{eqnarray}
Similarly, 
let $P_{\rm d}^{+}$ denotes
the error rate for the successive $\pi/2$ pulse controls that intend to make
$\theta=\pi/2\rightarrow3\pi/4$ ($\pi/2$ pulse) and $\theta=3\pi/4\rightarrow \pi$ ($\pi/2$ pulse).
We find that
\begin{eqnarray}
P_{\rm d}^{+}&\sim& P(\pi/2,\pi/4)+P(3\pi/4,\pi/4)\nonumber\\
&=&P^{+}+P^{-}\nonumber\\
&\sim&\frac{\pi^2+4}{32N}\bottom{.} 
\label{eq:Accum2}
\end{eqnarray}
(\ref{eq:Accum1}) and (\ref{eq:Accum2}) show that the error rates in the successive
$\pi/2$ pulse controls coincide with the semiclassical result 
$ \frac{\pi^2+4}{64N}\times 2$.
This may be understood as that the total controls in both cases ($\ket{0} \leftrightarrow \ket{1}$) can be regarded as classical flipping and thus the back action from the qubit becomes
invisible.
With the exception of  these special cases, 
successive control errors cannot be explained by 
semiclassical analysis.
For example, the error rates of the successive $\pi/2$ pulse controls starting from $\ket{\pi/4}$ and $\ket{3\pi/4}$,
denoted respectively
as $P_{\rm d}^{q}$ and $P_{\rm d}^{3q}$,  have different values as
\begin{eqnarray}
P_{\rm d}^{q}&\sim& P(\pi/4,\pi/4)+P(\pi/2,\pi/4)\sim\frac{(\pi+2)^2}{32N}\bottom{,} 
\\
\label{eq:Accum3}
P_{\rm d}^{3q}&\sim& P(3\pi/4,\pi/4)+P(0,\pi/4)\sim\frac{(\pi-2)^2}{32N}\bottom{.} 
\label{eq:Accum4}
\end{eqnarray}
We can also calculate the error rate for the $\pi$ pulse from  (\ref{eq:Ptheta}) as follows.
By setting $\theta=\pi/2$, (\ref{eq:Ptheta})  is rewritten for $\pi$ pulse control for
the initial state as
\begin{eqnarray}
 P(\theta_0,\pi/2) &=& \frac{1}{4}\sum_{n=0}^{\infty}|C_{n}|^2
\{m^{00}+ m^{11} +(m^{01}+ m^{10})\sin{2\theta_0}
\nonumber\\
&& +(m^{00}-m^{11})\cos2\theta_0\}^2 \bottom{.} 
\label{eq:PthetaPi}
\end{eqnarray}
Putting the elements $m_n^{ij}$ in (\ref{Eq:Defm}),  Eq.~(\ref{eq:PthetaPi}) for
the specific values of 
$\theta_0=0, \pi/2, \pm\pi/4$ becomes
\begin{eqnarray}
 P(0,\pi/2) &=& \sum_{n=0}^{\infty}|C_{n}|^2 \cos^2 (\kappa\sqrt{n})
\label{eq:PthetaPi1}
\bottom{,} 
\\
 P(\pi/2,\pi/2) &=& \sum_{n=0}^{\infty}|C_{n}|^2 \cos^2(\kappa\sqrt{n+1}) \bottom{,} 
\\
 P(\pm\pi/4,\pi/2) &=&\frac{1}{4}\sum_{n=0}^{\infty}|C_{n}|^2 
\big\{\cos (\kappa\sqrt{n})+ \cos (\kappa\sqrt{n+1}) \nonumber\\
&&\pm\sqrt{N/(n+1)}\sin (\kappa\sqrt{n+1})  \nonumber\\
&&\mp\sqrt{n/N}\sin (\kappa\sqrt{n})\big\}^2
\nonumber\\
&=&  \frac{1}{2}+\frac{1}{2}\sum_{n=0}^{\infty}|C_{n}|^2 
\bigg\{\cos (\kappa\sqrt{n+1}) \cos (\kappa\sqrt{n}) \nonumber\\
&&-\sqrt{\frac{n}{1+n}}
\sin(\kappa\sqrt{n+1}) \sin(\kappa\sqrt{n})\nonumber\\
&&
\mp\sqrt{\frac{n}{N}}\sin (\kappa\sqrt{n})
\big\{\cos(\kappa\sqrt{n+1})\nonumber\\
&&-\cos(\kappa\sqrt{n-1}) \big\}
\bigg\}\bottom{.} 
\label{eq:PthetaPi3}
\end{eqnarray}
Using $x\equiv(n-N)/N$ again and setting $\vartheta\equiv\kappa\sqrt{N}=\pi/2+O(1/N)$, 
we obtain the expansions  
for the summands in (\ref{eq:PthetaPi1}-\ref{eq:PthetaPi3}) as
\begin{eqnarray}
\cos^2 (\kappa\sqrt{n})&=&
\frac{1}{2}\Big\{1+\cos\Big(\pi\sqrt{1+x}\Big)\Big\}\nonumber\\
&\sim&\frac{\pi^2 }{16} x^2 \bottom{,} 
\label{eq:Expand1}
\end{eqnarray}
\begin{eqnarray}
\cos^2 (\kappa\sqrt{n+1})&=&
\frac{1}{2}\Bigg\{1+\cos\Bigg(\pi\sqrt{1+x+\frac{1}{N}}\Bigg)\Bigg\}\nonumber\\
&\sim&\frac{\pi^2 }{16} x^2
+\frac{\pi^2 }{32N}  (x-3x^2)
\bottom{,} 
\label{eq:Expand2}
\end{eqnarray}
\begin{eqnarray}
&&\cos (\kappa\sqrt{n+1}) \cos (\kappa\sqrt{n})
-\sqrt{\frac{n}{1+n}}\sin(\kappa\sqrt{n+1}) \sin(\kappa\sqrt{n})\nonumber\\
&=&
\cos\Bigg(\frac{\pi}{2}\sqrt{1+x+\frac{1}{N}}\Bigg)
\cos\Big(\frac{\pi}{2}\sqrt{1+x}\Big)\nonumber\\
&&-\sqrt{\frac{1+x}{1+x+\frac{1}{N}}}
\sin\Bigg(\frac{\pi}{2}\sqrt{1+x+\frac{1}{N}}\Bigg)
\sin\Big(\frac{\pi}{2}\sqrt{1+x}\Big)
\nonumber\\
&\sim&-1+\frac{1}{2N}+\frac{\pi^2 }{8} x^{2}
+\frac{\pi^2(x-2x^2)}{8N}
\bottom{,} 
\label{eq:Expand3}
\end{eqnarray}
\begin{eqnarray}
&&
\sqrt{\frac{n}{N}}\sin (\kappa\sqrt{n})
\big\{\cos(\kappa\sqrt{n+1})-\cos(\kappa\sqrt{n-1}) \big\}
\nonumber\\
&=& \sqrt{1+x}\sin\Big(\frac{\pi}{2}\sqrt{1+x}\Big)
\nonumber\\&&\cdot\Bigg\{\cos\Bigg(\frac{\pi}{2}\sqrt{1+x+\frac{1}{N}}\Bigg)
-\cos\Bigg(\frac{\pi}{2}\sqrt{1+x-\frac{1}{N}}\Bigg)\Bigg\}\nonumber\\
&\sim&-\frac{\pi}{2N}+\frac{\pi^3 x^2}{32N}
\bottom{,} 
\label{eq:Expand4}
\end{eqnarray}
where we have ignored  the terms of $O(x^3), O(xN^{-1}), O(N^{-2})$ 
in the approximations.
Eqs.~(\ref{eq:PthetaPi1})-(\ref{eq:PthetaPi3}) 
are carried out  in the same manner  as Eq.~(\ref{Eq:fsum})
with regarding the expansions (\ref{eq:Expand1})-(\ref{eq:Expand4})
as that of $f^{\pm}$ in (\ref{Eq:fexpansion}).
Thus, the error rates in the $\pi$ pulse controls up to the $1/N$ order
are obtained as
\begin{eqnarray}
 P(0,\pi/2) &\sim& 
\frac{\pi^2}{16N}\bottom{,}
\label{eq:PthetaPiValue1}\\
 P(\pi/2,\pi/2) &\sim& 
\frac{\pi^2}{16N}\bottom{,} 
\label{eq:PthetaPiValue2}\\
 P(\pm\pi/4,\pi/2) &\sim& 
\frac{(\pi\pm2)^2}{16N}\bottom{.} 
\label{eq:PthetaPiValue3}
\end{eqnarray}
The resulting error rates are listed in the Table~\ref{table2} and shown 
in the Figs.~\ref{fig:Graph0} and \ref{fig:Graph1}, which show that
the values of the error rates and error accumulation tendency differ
greatly depending on the initial $\theta_0$ of the qubit.

\begin{figure}[t]
\begin{center}
\includegraphics[scale=0.3]{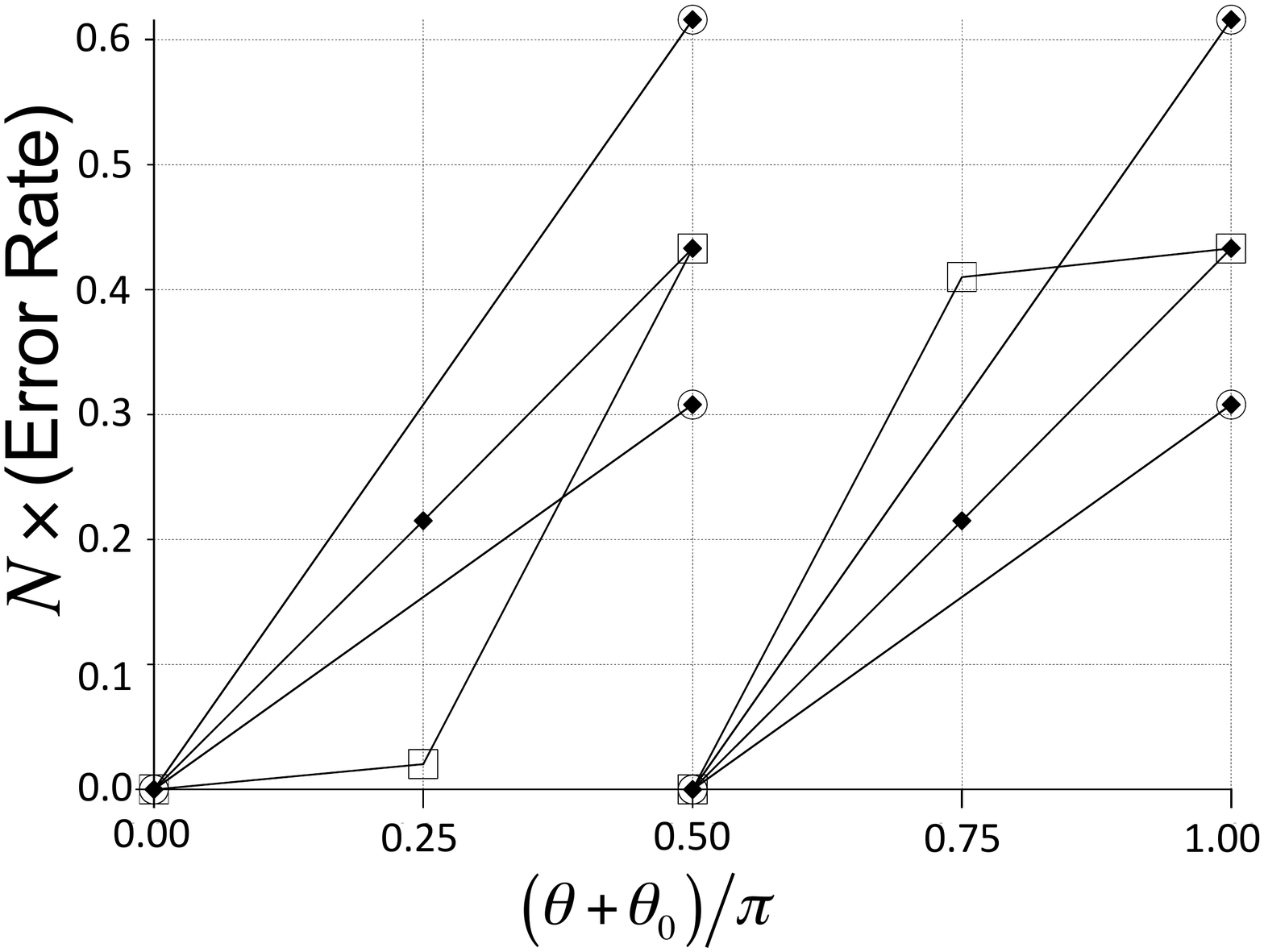}
\caption{
Error rates of the $\pi$ and $\pi/2$ controls($\theta=\pi/2, \pi/4$, respectively)  from the initial values $\theta_0=0$ and $\pi/2$
 are plotted for the target values $\theta_0+\theta$.
Open squares on the two kinked lines  represent the error rates for the successive $\pi/2$ pulse controls with
the average photon number $N$ in each control pulse, while
the open circles are for the single $\pi$ pulse controls with average photon numbers $N$(uppers)
and  $2N$(lowers),  all obtained  from the quantum analysis.
 The solid diamonds represent the corresponding error rates obtained from the classical analysis.
 The error rates obtained from 	quantum and classical analysis show good agreement except the kinked points.}
\label{fig:Graph0} 
\end{center}
\end{figure}

\begin{figure}[t]
\begin{center}
\includegraphics[scale=0.3]{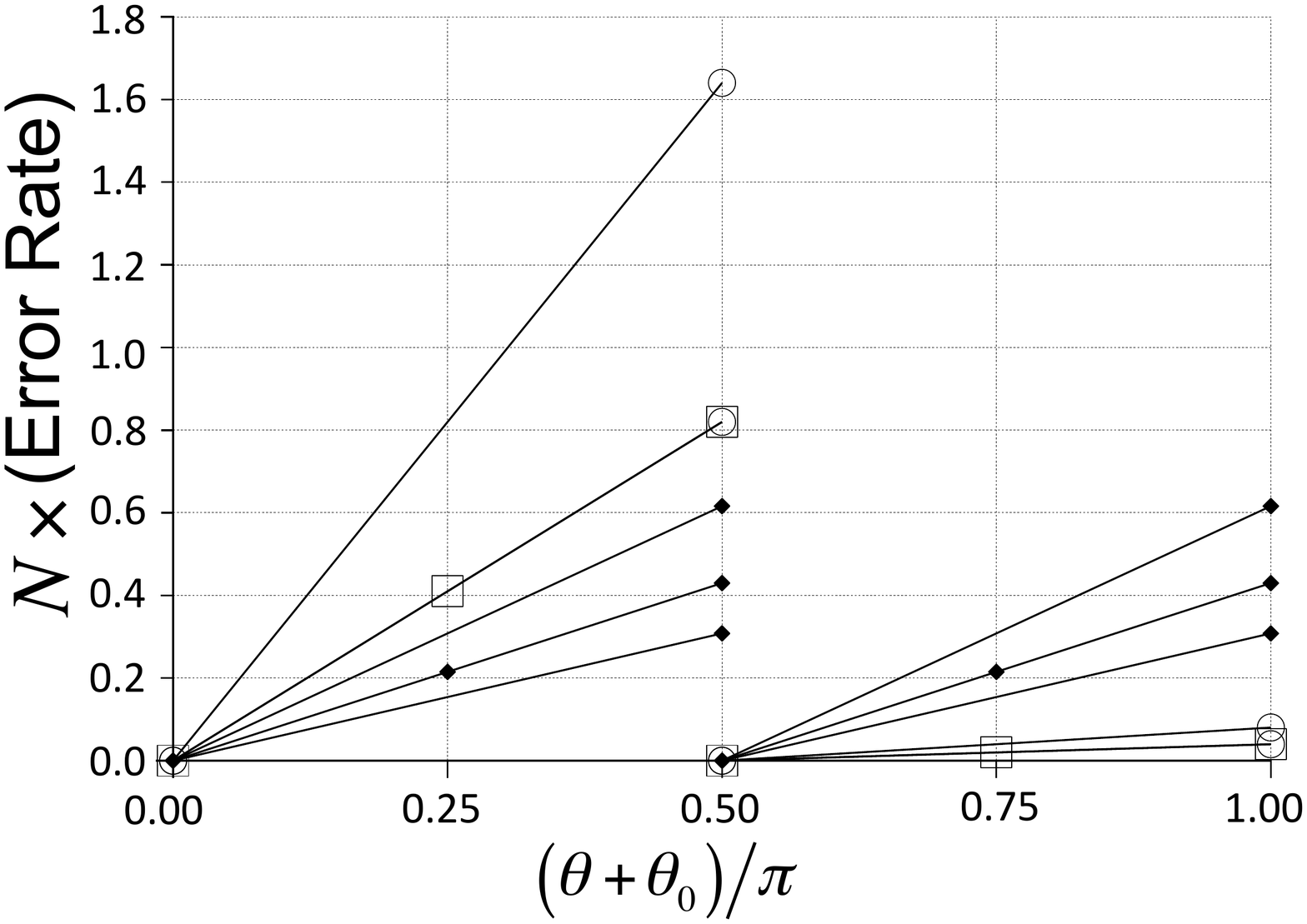}
\caption{
Error rates of the $\pi$ and $\pi/2$ controls($\theta=\pi/2, \pi/4$, respectively)  from the initial values $\theta_0=\pi/4$ and $3\pi/4$
 are plotted for the target values $\theta_0+\theta$.
 The legend is the same as used in  FIG.~2.
 In this case, successive $\pi/2$ pulses show linear error increase
 even with the quantum analysis.
 The quantum results are  totally different from the classical counterparts and
 show large discrepancy between the two initial conditions.
 }
\label{fig:Graph1} 
\end{center}
\end{figure}

\begin{table}
\begin{tabular}{|c|c|c|c|c|}\hline
$\theta\backslash\theta_0$&0&$\pi/4$&$\pi/2$&$3\pi/4$\\
\hline&&&&\\
$\pi/4$&\large$\frac{(\pi-2)^2}{64N}$&\large$\frac{(\pi+2)^2}{64N}$
&\large$\frac{(\pi+2)^2}{64N}$&\large$\frac{(\pi-2)^2}{64N}$
\\&&&&\\\hline&&&&\\
$\pi/2$
&\large$\frac{\pi^2}{16N}$
&\large$\frac{(\pi+2)^2}{16N}$
&\large$\frac{\pi^2}{16N}$&\large$\frac{(\pi-2)^2}{16N}$
\\&&&&\\\hline
\end{tabular}
\caption{Error rates $P(\theta_0,\theta)$ of $\pi/2$ and $\pi$ pulses for various initial states up to the order of $N^{-1}$. Effect of the qubit back-action appears in all the cases except those of $\pi$ pulse with $\theta_0=0,\pi/2$, which corresponds to the classical bit flip.}
 \label{table2} 
\end{table}
\section{Conclusions and discussion}
We have analyzed the dynamics of a qubit state controlled 
by coherent electromagnetic fields with a fully quantum 
Jaynes-Cummings interaction.
Although the control is assumed to be free from technical imperfections such as frequency detuning 
and interaction time fluctuation,  the control error  
inherent in the finiteness of the control fields is found to limit the precision of the control.
The error rate in $\pi/2$ pulse control with the coherent state of the average photon number $N$  has turned  out to be $\frac{(\pi-2)^2}{64N}$ by the measure defined by (\ref{eq:Fidelity}) when the qubit is initially in the ground state.
It is about 9\% of the previously reported error rate, $\frac{\pi^2 +4}{64N}$ obtained from a semi-classical model~\cite{GB}.
Since this control can be used  to prepare the initial state of the qubit register,
a one order of magnitude increase in the accuracy (or 90 \% energy saving) 
is  good news.
When the initial state of the qubit is 
the excited state, the error rate in the same control becomes $\frac{(\pi+2)^2}{64N}$. 
The semi-classical model underestimates the error rate by
1.9 times in this case.

We also have shown  the error accumulation in controls using two successive $\pi/2$ pulses.
In both cases of a qubit starting from the ground state and from the excited state, 
the overall error rate coincides with
 the semi-classical result as $2\times\frac{\pi^2 +4}{64N}$ except that the accumulation tendency is not linear, as shown in Fig.~\ref{fig:Graph0}.  It is also shown that 
 the total error rate is smaller than the single
$\pi$ pulse control with a coherent field containing $N$ photons on average,
but is larger than the $\pi$ pulse control with a field with $2N$ photons,
which uses the same total energy.

As shown in Fig.~\ref{fig:Graph1}, the error accumulation of the successive $\pi/2$ pulses are linear in the cases where the initial qubit state vector lies in the XY plane in the
Bloch sphere.  The successive $\pi/2$ pulses yield the same error rate with the single
$\pi$ pulse  if 
the total energies in the control fields are the  same.
But the error rates can differ by as much as  20 times,
depending on the initial state of the qubits.
This suggests that the error rate takes its minimum and maximum values when the midpoint of the rotation of the Bloch vector is located at the bottom and the top of the sphere,
respectively. This conjecture is readily confirmed by calculating the general error rate  $P(\theta_0,\theta)$ 
in (\ref{eq:Ptheta}) up to the order $1/N$, which can be performed in the same manner as was done for  $P(0,\pi/4)$ and $P(\pi/2,\pi/4)$ in Sec.\ III and IV.
By using the altitude angle $\Phi$ of the midpoint of the rotation 
instead of $\theta_0$, the result is expressed as
\begin{equation}
P\left(\Phi-\frac{\theta}{2}, \theta\right)\sim\frac{(\theta-\cos{2\Phi}\sin\theta)^2}{4N}\bottom{.} 
\label{eq:General}
\end{equation}
Eq.~(\ref{eq:General})  gives $\Phi$ values that provides the maximum(minimum) value
of the error rate at $\pi/2$(0) for the domain $0\leq\theta\leq\pi$.
The maximum and minimum error rates in the $\theta/2$ pulse control 
are written as
\begin{equation}
P_{\rm max}(\theta)\sim\frac{(\theta+\sin\theta)^2}{4N}
\label{eq:MaxError}
\end{equation}
and
\begin{equation}
P_{\rm min}(\theta)\sim\frac{(\theta-\sin\theta)^2}{4N}\bottom{.} 
\label{eq:MinError}
\end{equation}
The value of $P_{\rm max}(\theta)$  is important because it is
used to estimate the maximum error 
during quantum computations where qubit states are unknown before control.

To conclude the paper, we discuss the validity of the pulse area theorem.
We have proven the theorem  rigorously up to the order $1/N$.
If the  deviation of $\vartheta$ from $\pi/4$ 
is in the order of $N^{-1}$,
 the error becomes quantum-mechanical limited
and solely stems from the finiteness of the control field.
If we consider the error rate of the order of  $N^{-2}$, the minimum value will be
attained at $\vartheta^{\pm}$ derived in (\ref{Eq:thetaopt}).

\section{ACKNOWLEDGMENTS}
K.~I. thanks to Y.~Tokura for fruitful discussions
and useful suggestions.
This work was supported by the
MEXT Grants-in-Aid for Scientific Research on Innovative Areas
20104003 and 21102008.

\end{document}